\documentstyle[preprint,aps]{revtex}
\begin{document}
\title{
Numerical Results for the Ground-State Interface in a Random Medium
}
\author{
A. Alan Middleton
}
\address{
Department of Physics, Syracuse University, Syracuse, NY 13244
}
\date{\today}
\maketitle
\begin{abstract}
The problem of determining the ground state of a
$d$-dimensional interface embedded in a
$(d+1)$-dimensional random medium is treated numerically.
Using a minimum-cut algorithm,
the exact ground states can be found for a number of problems
for which other numerical methods are inexact and slow.
In particular, results are presented for the roughness exponents
and ground-state energy fluctuations in a random bond Ising
model.
It is found that the roughness exponent
$\zeta =  0.41 \pm  0.01,
 0.22 \pm  0.01$, with the related energy exponent
being $\theta =  0.84 \pm  0.03,
 1.45 \pm  0.04$,
in $d = 2, 3$, respectively.
These results are compared with previous analytical and
numerical estimates.

\end{abstract}



In many physical systems, the equilibrium and non-equilibrium
behavior of interfaces determine the thermodynamic and dynamic
properties of the system.
Examples of such interfaces include the phase
boundary between spin-up and spin-down domains in an Ising
magnet and the fluid-vapor interface in porous media.
A number of problems in statistical mechanics can be mapped
onto questions about surfaces in a higher dimensional
system; for example, the Potts models
can be cast as ``height models''
\cite{heightmodel}.
The study of interfaces has also been applied to the
fracture in solids, where the fracture location relative to
the original whole body is treated as an interface
\cite{KardarFracture}.
Quenched disorder greatly affects the equilibrium
and dynamical properties of interfaces and phase boundaries
\cite{manyrefs}.
An interface is distorted by the
random forces or potential caused by the disorder, so that
its shape is much rougher than in a pure
system.

The problem considered here is that of finding the
ground state, or zero-temperature configuration,
of a $d$-dimensional interface in a $(d+1)$-dimensional medium
with quenched disorder, i.e., spatially varying couplings that
are fixed in time.
Given a method for finding
the minimal energy configuration of an interface for a
particular realization of the disorder, statistical properties
can be found by averaging over many realizations.
One can then determine, for example,
how the perpendicular width $W$ of the interface and
the fluctuations
in the ground state energy depend on the linear size $L$ of the
interface.
These quantities capture many of the features of
the low energy state and
are relevant to thermodynamic properties of the interface at
low temperatures.
This information can also be used to place lower bounds on
the barriers between low energy configurations
\cite{manyrefs,barriers}, to study the
sensitivity of the location of the interface to
perturbations \cite{sensitivity}, and
to help to analyze the motion of an interface subject to an
external drive \cite{barriers}.

The approach to finding the ground state used in this paper
applies a minimum-cut (or ``mincut'') algorithm \cite{mincutref}
to find the exact ground state.
Algorithms for finding the mincut take a time bounded
by a polynomial in the
volume of the system;  finding the ``best'' algorithm is still a
matter of active research in combinatorial optimization.
The polynomial time bound means, however, that this algorithm is
extremely fast compared to annealing techniques, exhaustive
search techniques, or transfer-matrix
algorithms \cite{kardarzhangtransfer}
that take a time exponential in the size of the system.
Given such an algorithm, quantities of interest can be measured
and averaged over many
realizations of the quenched disorder.
A type of mincut algorithm has been applied previously to the
random field Ising magnet \cite{ogielski}.
The mincut algorithm is applicable to many other
physical problems, however.
In this paper, I describe the general technique and
summarize the results for the interface in the random bond
Ising model.

The problem of finding the ground state of a random bond
Ising model (RBIM) in zero field is that of minimizing the total
energy
 \begin{equation}
E = -\sum_{\left<ij\right>} J_{ij} s_i s_j,
 \end{equation}
where the spin variables $s_i$ take on the values $\pm 1$
and the sum is over nearest neighbor pairs $\left<ij\right>$ on a
$(d+1)$-dimensional lattice.  The $J_{ij} \geq 0$ are independent
identically-distributed random variables; in this paper, the
$J_{ij}$ are taken to be uniformly distributed in
the range $\left[0 ,  1 \right)$.
I consider here a lattice of dimensions $H \times L^d$, with
coordinates $\{y, x_1, \ldots, x_d\}$, $0 \leq y < H$, and
$0 \leq x_\alpha < L$ for $1 \leq \alpha \leq d$.
Boundary conditions must be chosen
to induce an interface;
this is done by setting the spins $s_i$ to have the value $+1$
on the surface $y = 0$ and to be equal
to $-1$ on the surface $y = H-1$.
Nearest neighbors are defined so that the constant $x_0$ surfaces
represent slices of a ``cubic'' lattice taken along the $[11 \ldots 1]$
direction \cite{slicenote}.
The RBIM is used to describe a ferromagnet with random
variation in the coupling strength due to spatial disorder of the spins
or impurities.
This model has also been used to describe fracture
in materials, where the $J_{ij}$ represents the local force needed
to break the material \cite{KardarFracture}
and it is assumed the fracture
occurs along the surface of minimum total rupture force.

The problem of the random bond Ising magnet was treated numerically
by Huse and Henley \cite{husehenley} for $d=1$, in the
case of the directed polymer, where
the interface is restricted
to be described by a single valued function $y(x_1)$.
The algorithm used, an iterative transfer-matrix method,
will find the lowest energy state subject to this restriction.
Though the algorithm used here need not be subject to this
restriction, the single-valued-$y$ approximation
is convenient for analytical work
and is believed to not affect the long wavelength behavior.
Huse and Henley found that the transverse width
 \begin{equation}
W = \{\sum_{\{x_\alpha\}} y^2(x_\alpha) -
     [\sum_{\{x_\alpha\}} y(x_\alpha)]^2\} ^ {1/2}
 \end{equation}
scaled with the length $L$ of the interface as
$W^\zeta$, with $\zeta \approx 0.66$.  It was also found that
the sample-to-sample rms fluctuations in the ground state energy
varied as $\Delta E \sim L^\theta$, with $\theta \approx 0.33$.
Subsequently the $d=1$ case was solved analytically by Huse, Henley
and Fisher \cite{husehenley}, who showed that $\zeta = 2/3$ and
$\chi = 1/3$ in $d=1$, consistent with the numerical
work of Huse and Henley.
Numerical work for interface dimension $d=2$
was done by Kardar and Zhang
\cite{kardarzhangtransfer},
using a transfer matrix which operates
on the space of paths; this algorithm, though exact, requires a
time exponential in one of the dimensions of the system; they
were consequently restricted to lattices with linear
dimension $L\leq 16$.
{}From their simulations, they
found $\zeta = 0.50 \pm 0.08$ and $\theta = 1.10 \pm 0.05$
in $d=2$.
Analytical arguments have been given for the roughness exponents
in higher dimensions
\cite{KardarFracture,FisherRBfuncrenorm,HalpinHealy}.

The algorithm used to find the ground state is based upon a
maximum network flow algorithm.
The network flow problem is defined
on a graph with given ``capacities'' indicating the
rate at which fluid can flow from one node of the graph
to another along a directed edge that connects the two nodes.
The goal is to determine
the maximum amount of flow that can be sustained
between two given points, the source and the sink,
given current conservation at each of the non-terminal
nodes.  It is straightforward to show that the maximum
flow value is equivalent to
the value for the minimum ``cut'', which is defined as the weight
of the set of edges with
minimum total capacity that, when cut, disconnect the source and
the sink.  This minimum cut can be imagined
as describing a minimal
flow surface that separates the source
and the sink by intercepting
the flows through the removed edges.  The equivalence of the
network flow and the minimum cut is
due to the fact that the minimum
cut is the ``bottleneck'' through which all flow must pass.

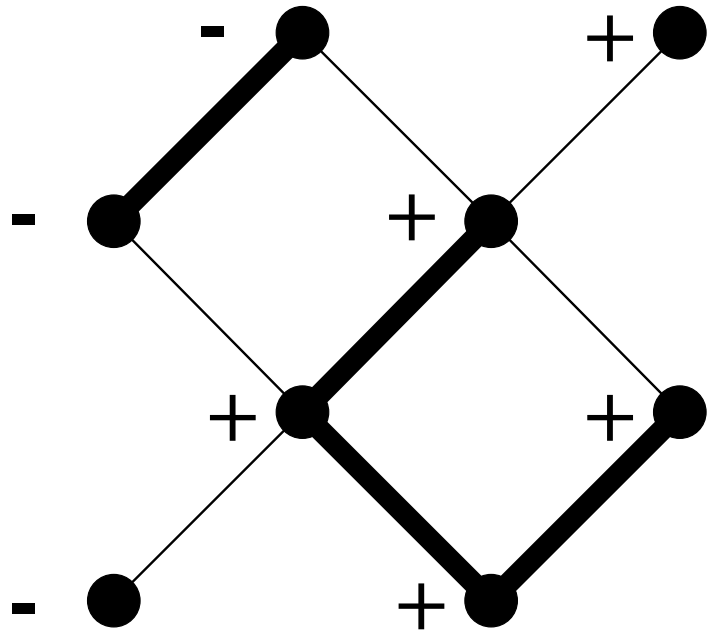
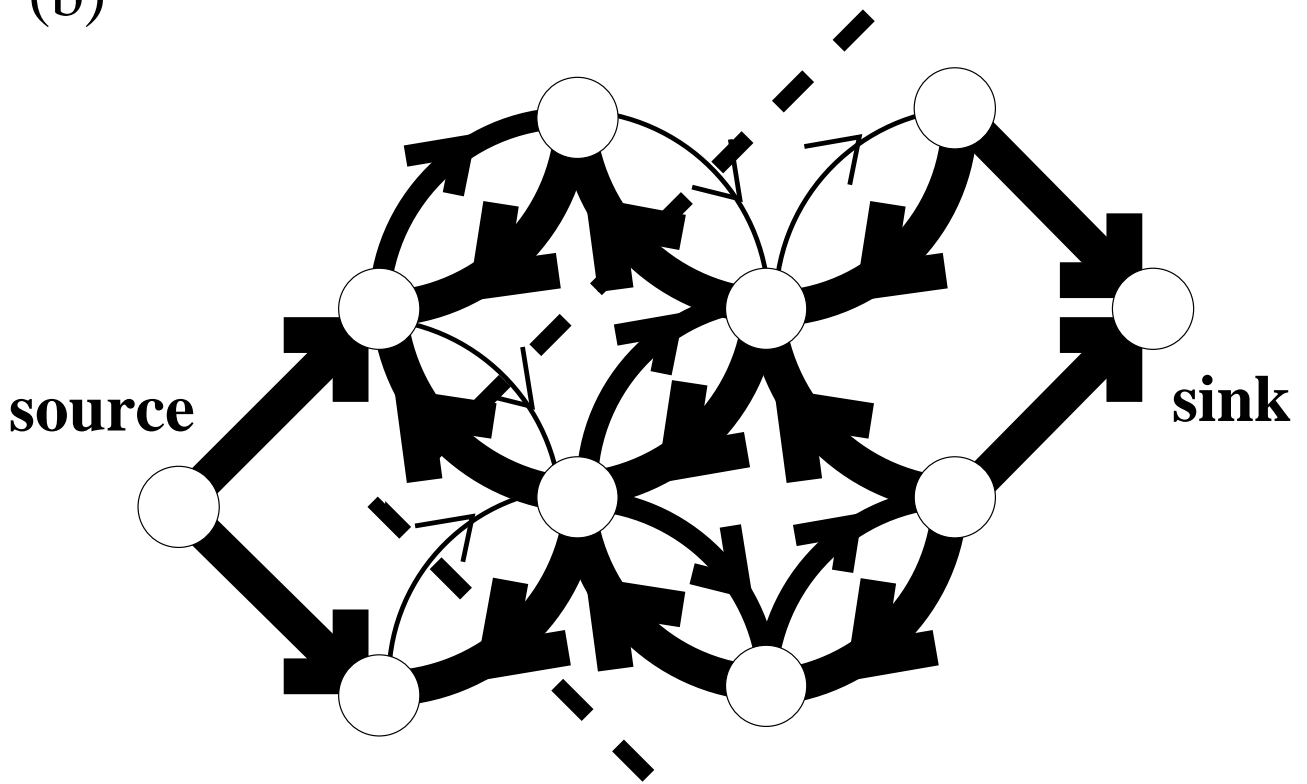
\begin{figure}
\caption{
Example of the construction of the graph whose minimum-cut
corresponds to the ground state of a
random-bond Ising model with twisted boundary conditions.
(a) The sites and bonds for an Ising problem, with
dimensions $L=2$ and $H=4$ (in $(d=1)+1$ dimensions).
The stronger bonds (larger $J_{ij}$)
are indicated by thicker widths.  The sites are labeled
according to their values in the ground state configuration,
with sites fixed to take on the values $-$ and $+$ on the
left and right sides, respectively.
(b)  The corresponding graph for the minimum-cut problem.
The ``source'' and ``sink'' nodes are auxiliary nodes added to
the bulk lattice to define the boundary conditions.
Nodes are connected by directed edges, with
the capacity of the edges indicated
by the width of the lines.
The bulk forward edges have capacities $J_{ij}$.
``Forbidden'' backwards
edges with large weights (greatest thicknesses)
both enforce boundary conditions and prevent overhangs in this
example; see text for the case with overhangs.
The minimum-cut surface separating the sink from the source
is indicated by the broken line.
}
\label{graphmap}
\end{figure}

To apply this algorithm to the RBIM, a graph is constructed whose
minimum cut directly reflects the minimum energy interface in the
spin lattice.
This is done by attaching two fictitious spins
to the lattice, the source and the sink,
which can be seen as setting
the boundary conditions for the spin in the Ising magnet.
The source is connected by an
edge to each $y=0$ site; each $y=H-1$
site is connected to the sink.  The capacity on each of these
auxiliary edges is set
to be a very large value ($2d$ times the
maximum $J_{ij}$) so that the
minimum cut surface does not pass
through these edges and instead lies in the bulk.
One then has the choice of whether or not
to implement the single-valued-$y$
approximation, according to the choice of the capacities of
the edges connecting the bulk nodes.  If for each pair of nearest
neighbors $\left< i j \right>$, two edges are created, one from
$i$ to $j$ and the other from $j$ to $i$, each with capacity
$2 J_{ij}$, the maximum flow algorithm
will give the interface and energy of the unrestricted interface
problem.  The cost of the minimum cut, i.e., the
minimum of $2 \sum J_{ij}$ over all surfaces which
separate the source
and the sink, is then equal to the minimal energy of an
interface, above the ground state (which has all spins equal).
See Fig.\ (\ref{graphmap}) for a diagram of this construction.

For $d=2,3$, I simulated the restricted, single-valued,
problem by choosing the capacity of the edge connecting $i$ to $j$
to be $J_{ij}$ if $y(i)<y(j)$ and to be $2d\max(J_{ij})$ if
$y(i) > y(j)$. (Because of the choice of lattice alignment,
$y(i) \ne y(j)$ for nearest neighbors $\left< i j \right>$.)
The high cost of the ``backwards'' edges, those connecting higher
$y$-coordinate sites to those with smaller $y$-coordinates, excludes
them from the minimal cut \cite{needback}.
In order to minimize the effects of the boundary surfaces on the
energy associated with the minimum cut, the edge costs were increased
to $2d\max(J_{ij})$ on a single line of bonds from one boundary to
the other, except on the center edge.
The minimal cut surface will not intersect this line except
at the midpoint, where the energy is not raised.
This modification therefore ``pins'' the interface at one point,
which is halfway between the boundaries.

Once the corresponding graph is constructed, the max-flow/min-cut
algorithm can be directly applied.
Specifically, the algorithm and code developed by Cherkassky
and Goldberg \cite{CherkasskyGoldberg} was adapted to the
class of graphs studied here;
this code was approximately 2-3 times faster than other
code applied to this problem.
The computer time for the solution of a particular realization
on an IBM RS/6000-390 workstation
was found to average 230s for $d=2, L=120, H=50$ and
146s for $d=3, L=30, H=20$, with the time for smaller systems
being nearly proportional to the volume of the sample.
The program gives both the minimum-cut cost, i.e., the ground state
energy, and the location of the minimum-cut.
The width of the interface was computed by finding the rms fluctuation
of the $y$-coordinate of the cut bonds; the cut can be visualized
by drawing the plaquettes that are dual to the cut bonds.

The results for the sample-averaged interface widths and energy
fluctuations are shown in  Fig.\ (\ref{figresults2d}) and  Fig.\
(\ref{figresults3d}),
for $d=2$ and $d=3$, respectively.
The scaling plots assume a scaling form for the average width
 \begin{equation}
\overline{W(L, H)} = L^{\zeta} w( H / L^{\zeta}),
 \end{equation}
with $w(x) \sim x$ for $x \rightarrow 0$ and
$w(x) \sim const$ for $x \rightarrow \infty$, and a scaling form for the
rms energy fluctuations of
 \begin{equation}
\overline{\Delta E(L, H)} = L^{\theta} u( H / L^{\zeta}),
 \end{equation}
with $u(x) \sim x^{\theta/\zeta}$ for $x \rightarrow 0$ and
$w(x) \sim c$ for $x \rightarrow \infty$.
The best scaling fits for the largest $L$ are obtained for the
values
$\zeta =  0.41 \pm  0.01,
 0.22 \pm  0.01$ and
$\theta =  0.84 \pm  0.03,
 1.45 \pm  0.04$,
in $d = 2, 3$, respectively.
The statistical errors due to sample-to-sample
fluctatuations in the width $W$
are relatively small (for almost all sizes, $10^4$ samples
were generated; at least $10^3$ samples were generated in all
cases).
The statistical errors in the energy fluctuations $\Delta E$ are
somewhat larger, as can be seen from the scatter in the plots.
The dominant source
of errors in these fits are from the unknown corrections to scaling;
the uncertainty in the exponents indicates the range of values over
which the curves for the three largest $L$ values agree to within
statistical error when scaled.
The larger relative errors in $\theta$ are also due to a
long crossover in
the scaling plots;
the widths $W$ appear to converge more
quickly, yet also have a slow convergence after a sharp crossover.
The exponent values satisfy the scaling relation
$2\zeta = \theta + d - 2$
to within numerical error \cite{husehenley}.
Applying this scaling relation using the numerical values
of $\zeta$ gives
$\theta =  0.82 \pm  0.02,
 1.44 \pm  0.02$ for $d=2,3$, respectively.
The exponents found here are in disagreement with the results of
Ref.\ \cite{kardarzhangtransfer}; this is likeley due to
finite size effects in the smaller systems examined there.
Given the current numerical accuracy, it is difficult to differentiate
between the analytical results of Fisher and those of
Halpin-Healy, which are very
close in value ($\zeta = 0.208 (4 - d)$
and $\zeta = 2(4-d)/9$, respectively).
The numerical results are
in agreement with both.
The constraint preventing computations for
larger systems on workstations is
memory ( $> 256$ MB will be required)
rather than total CPU time.

The max-flow algorithm for determining min-cuts has been applied
to several problems in the past, most notably the random field
Ising magnet \cite{ogielski}, but this
is the first time that
it has been applied directly to the random bond Ising model.
As has been noted, this technique is also applicable to other
systems of current interest.
This technique is not restricted to single-valued height
interfaces.
For example, from simulations without the
single-valuedness restriction,
I find the roughness exponent $\zeta = 0.67 \pm 0.02$
for $d = 1$,
in agreement with theory and numerical results for the case
where overhangs are forbidden (the directed polymer problem)
\cite{manyrefs}.

I am pleased to acknowledge discussions with S. Rao,
M. Kardar, and Y. Shapir.
The majority of the
computational work was conducted using the resources of
the Northeast Parallel Architectures Center (NPAC) at Syracuse
University.

\begin{figure}
\caption{
Scaling plots for average ground state quantities in $2+1$ dimensions.
All lengths are measured in units of the lattice constant.
(a) Scaling of the average sample width $W(L, H)
= \overline{<y^2(x)> - <y(x)>^2}$.
The scaled width $W/L^\zeta$ is plotted vs.\ the
scaled transverse sample size $H/L^\zeta$ for
different $L$, using the value
$\zeta =  0.41 \pm  0.01$.
(b) Scaling plot for the sample-to-sample fluctuations in the
ground-state energy.
The scale energy fluctuations $\Delta E/L^\theta$ are plotted
as a function
of scaled transverse sample size $H/L^\zeta$ for
varying $L$, using
$\theta =  0.84 \pm  0.03$.
}
\label{figresults2d}
\end{figure}

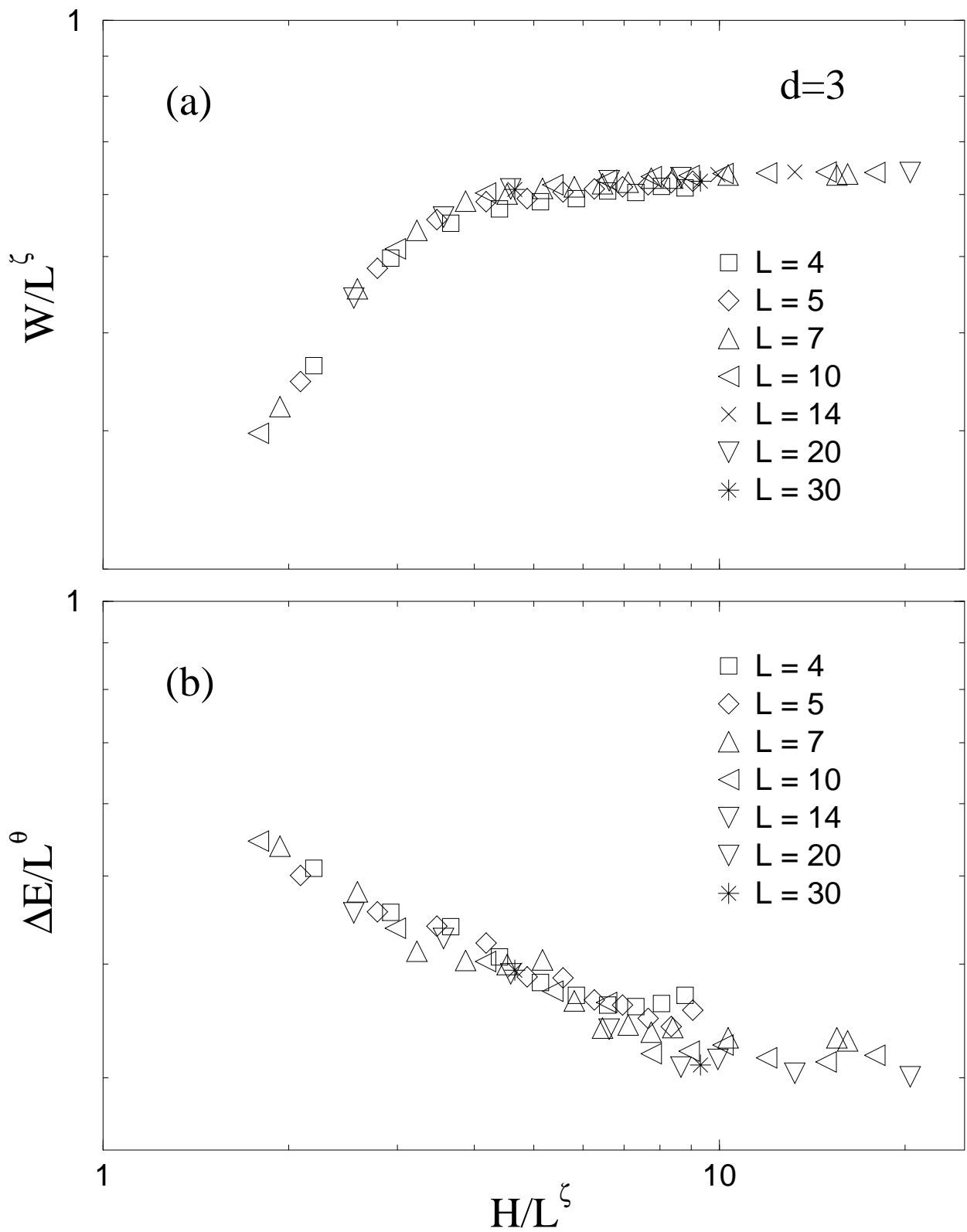
\begin{figure}
\caption{
Scaling plots for average ground state quantities in $3+1$ dimensions.
The plotted quantities are the same as in  Fig.\ (\ref{figresults2d}), with
$\zeta =  0.22 \pm  0.01$ and
$\theta =  1.45 \pm  0.04$.
}
\label{figresults3d}
\end{figure}


\begin{references}
\bibitem{slicenote}
This choice of constant $x_0$ surfaces (or, equivalently, boundary
conditions) lowers the
elasticity of the interface relative to the case where
the boundaries are perpendicular to the $\left[00\ldots 1\right]$
direction; the former choice reduces lattice effects
and is generally used for the directed polymer
problem in two dimensions.
The neighbors of the site at $\{y, x_1, \ldots, x_d\}$ have
coordinates $\{y+1, x_1, \ldots, x_d\}$,
$\{y+1, x_1+1, \ldots, x_d\}$, \ldots, $\{y+1, x_1, \ldots, x_d+1\}$.
\bibitem{needback}
If the ``backwards'' edges are not included in the
graph at all, one gets overhangs of
an unusual type, as backwards facing edges will not have any cost.
\end{references}
\end{document}